\documentclass[%
 reprint,
 amsmath,amssymb,
 aps,
]{revtex4-2}

\usepackage{graphicx}
\usepackage{dcolumn}
\usepackage{bm}
\usepackage{braket}
\usepackage{physics}
\usepackage{siunitx}
\DeclareSIUnit{\sample}{S}
\usepackage{color}

\usepackage[normalem]{ulem}

\newcommand{\rhothermal}{\hat{\rho}_{\textrm{th}}}
\newcommand{\nbarthermal}{n_{\textrm{th}}}

\newcommand{\omegac}{\omega_\text{c}}
\newcommand{\omegam}{\omega_{\textrm{m}}}

\newcommand{\Opd}[2]{\mbox{$\hat{#1}_{#2}^{\dagger}$}}
\newcommand{\Op}[2]{\mbox{$\hat{#1}_{#2}$}}

\newcommand{\kappae}{\kappa_{\textrm{e}}}
\newcommand{\kappai}{\kappa_{\textrm{i}}}

\newcommand{\aout}{\hat{a}_{\textrm{out}}}
\newcommand{\Aout}{\hat{A}_{\textrm{out}}}

\newcommand{\ain}{\hat{a}_{\textrm{in}}}
\newcommand{\aini}{\hat{a}_{\textrm{in,i}}}

\newcommand{\bin}{\hat{b}_{\textrm{in}}}
\newcommand{\aopd}{\hat{a}^\dagger}

\newcommand{\aop}{\hat{a}}

\newcommand{\R}{\Op{R}{}{}}
\newcommand{\povmRS}{\Op{\Pi}{\textrm{RS}}(r)}
\newcommand{\povmalpha}{\Op{\Pi}{}(\alpha)}
\newcommand{\disp}{\Op{D}{}{}(\alpha)}
\newcommand{\dispdagger}{\Opd{D}{}{}(\alpha)}
\newcommand{\bopd}{\hat{b}^\dagger}
\newcommand{\bop}{\hat{b}}

\newcommand{\gammai}{\gamma_{\textrm{i}}}

\newcommand{\gammaOM}{\gamma_{\textrm{OM}}}

\newcommand{\etaf}{\eta_{\textrm{f}}}
\newcommand{\etac}{\eta_{\textrm{c}}}

\newcommand{\alphaLO}{\alpha_{\textrm{LO}}}

\newcommand{\ddt}{\frac{d}{dt}}

\begin{document}

\preprint{APS/123-QED}

\title{Room-temperature Mechanical Resonator with a Single Added or Subtracted Phonon}

\author{Rishi N. Patel}
\email{rishipat@stanford.edu}
\author{Timothy P. McKenna}%
\author{Zhaoyou Wang}
\author{Jeremy D. Witmer}
\author{Wentao Jiang}
\author{Rapha\"{e}l Van Laer}
\author{Christopher J. Sarabalis}
\author{Amir H. Safavi-Naeini}
\email{safavi@stanford.edu}
\affiliation{Ginzton Laboratory, Stanford University Department of Applied Physics}%

\begin{abstract}

A room-temperature mechanical oscillator undergoes thermal Brownian motion with an amplitude much larger than the amplitude associated with a single phonon of excitation. This motion can be read out and manipulated using laser light using a cavity-optomechanical approach. By performing a strong quantum measurement, \emph{i.e.}, counting single photons in the sidebands imparted on a laser, we herald the addition and subtraction of single phonons on the $\SI{300}{\kelvin}$ thermal motional state of a $\SI{4}{\giga\hertz}$  mechanical oscillator. To understand the resulting mechanical state, we implement a tomography scheme and observe highly non-Gaussian phase-space distributions. Using a maximum likelihood method, we infer the density matrix of the oscillator and confirm the counter-intuitive doubling of the mean phonon number resulting from phonon addition and subtraction.
\end{abstract}

\maketitle

A mechanical oscillator at room temperature will behave in nearly perfect accordance with the laws of classical statistical physics. Nonetheless, interaction with optical-frequency photons can lead to behaviour that is nonclassical. For the mechanical motion of trapped ions, laser cooling and coupling of motion to the internal electronic degrees of freedom has long been pursued as a path to realizing a scalable quantum computer~\cite{cirac1995quantum}. For solid-state mechanical devices, the signatures of quantum noise and back-action have been observed at room temperature and proposed as a means to realize quantum sensors~\cite{purdy2017quantum,sudhir2017appearance}. A key feature of such optomechanical devices is their ability to efficiently generate correlations between motion and light~\cite{palomaki2013entangling}. A strong quantum measurement of the resulting light field, e.g, by using a single-photon detector can consequently alter the state of the mechanical system~\cite{cohen_phonon_2015,Riedinger2016,riedinger2018remote,enzian2021single,davis2018painting,Velez2019}. 

Experiments in nonlinear optics have shown that adding or subtracting photons fundamentally alters the state of an optical degree of freedom. For example, in photon addition, parametric down conversion followed by post-selection of an idler photon allows the preparation of a single-photon state from vacuum~\cite{Lvovsky2001}. Photon subtraction, when operating on squeezed light is used to generate and enlarge Schr{\"o}dinger's cat states~\cite{ourjoumtsev2006generating,sychev2017enlargement}. Both photon addition and subtraction are essential for a number of tasks in continuous variable quantum information processing, in which non-Gaussian states are often required~\cite{lvovsky2020production}. Highly thermal states of light, where significant classical noise would be expected to wash away any quantum effects, have also been converted to non-classical states by the addition of one photon~\cite{zavatta_experimental_2009,Kiesel,Kiesel2011,Vidrighin2016}. Such states, combined with photon subtraction, have also allowed a direct test of quantum commutation relations~\cite{Zavatta,Parigi2007}.
 
Mechanical oscillators have emerged as an important avenue for realizing quantum technologies. Cryogenically cooled oscillators are used more widely due to significantly reduced dissipation and reduction of thermal noise from the environment, which usually masks quantum features. Motivated by the long intrinsic relaxation times possible in cryogenically cooled mechanical oscillators~\cite{maccabe2019phononic}, and success in strongly coupling them to superconducting qubits~\cite{OConnell2010,Chu2018,Satzinger2018a,arrangoiz2019resolving}, proposals for realizing quantum machines that leverage their coherence and small size~\cite{pechal2018superconducting, hann2019hardware,chamberland2020building} have emerged recently. Separately, cavity-optomechanical approaches for quantum sensing and transduction are being pursued by groups around the world. Nonetheless, there is significant interest in operating quantum systems at higher temperatures since for many applications operation at ambient conditions is essential.

In this work, we perform experiments on a mechanical oscillator in a highly thermal state at room temperature, and use its interaction with optical photons to perform quantum-limited read-out and state control. First, we use single photon counting to prepare phonon-added and subtracted mechanical states in a regime where $kT\gg\hbar\omegam$. Second, building on tomography techniques in quantum optics~\cite{Lvovsky2009,lvovsky2020production,Lvovsky2001,eichler_experimental_2011} and optomechanics~\cite{vanner_cooling-by-measurement_2013,Muhonen2019}, we combine heralded single-phonon addition with the continuous measurement of mechanical amplitude and phase fluctuations. We reconstruct the density matrix of a mechanical oscillator in a phonon-added or phonon-subtracted thermal state with initial mean phonon occupancy of approximately $kT/\hbar\omegam = 1580$. With future improvements, our technique may be extended to prepare and characterize more complex states of mechanical motion.

\begin{figure}
\includegraphics{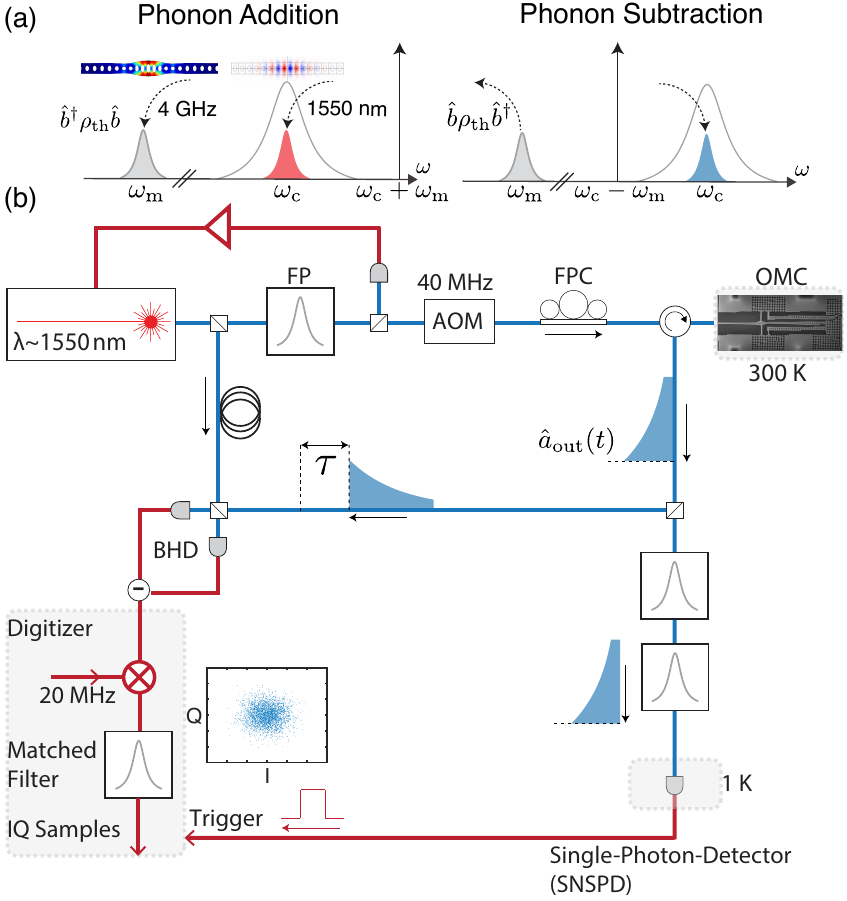}
\caption{\label{fig:setup} \textbf{Concept and principal components of the experimental setup} (a) Concept showing on left (right) how a Raman scattered photon on cavity-resonance at frequency $\omegac$ heralds the addition (subtraction) of a phonon at the mechanical frequency $\omegam$. Insets from left to right show simulated mechanical and optical mode profiles respectively. (b) Diagram showing the main components of the setup. A tunable laser is stabilized to a Fabry-Perot cavity (FP) and is frequency upshifted using an acousto-optic modulator (AOM). The resulting signal polarization is set using a fiber polarization controller (FPC) and sent into the optomechanical crystal device (OMC) whose temperature is stabilized near $\SI{300}{\kelvin}$. The reflected light is incident on a beamsplitter and is split into two paths. One (going to the bottom) contains pump-rejection filtering with a pair of FP filters, followed by single photon detection using a superconducting nanowire single photon detector (SNSPD). The other path (going to the left) performs balanced heterodyne detection (BHD) where the resulting signal is further digitally downconverted from $\SI{20}{\mega\hertz}$, and filtered using a digitizer and GPU. The digitizer captures $\approx \SI{2}{\micro\second}$ of data every time it is triggered by a single photon pulse from the SNSPD. A schematic of the decaying cavity field is shown inset in the figure. $\tau$ denotes the time-lag between when the click occurs and when the heterodyne data collection starts, as described in the text.}
\end{figure}

The process of phonon-addition and subtraction arises from the inelastic scattering of light from a laser due to mechanical motion in a cavity. By energy conservation, photons scattered by mechanical motion are shifted to a lower (higher) frequency corresponding to addition (subtraction) of a phonon in the mechanical resonator (Fig.~\ref{fig:setup}a). Selection of the desired process (addition or subtraction) is achieved by using the optical cavity resonance as a filter and tuning the laser to its blue or red side. This can be understood by considering the Hamiltonian that describes the optical and mechanical systems. For blue-detuning it is given by $\hat{H} = \hbar G (\aopd\bopd + h.c.) $ where $G$ is the linearized optomechanical coupling rate, and the annihilation operators for optical and mechanical oscillators are given by $\aop$ and $\bop$ respectively. Similarly, for red-detuning we have $\hat{H} = \hbar G (\aopd\bop + h.c.)$ whereby a phonon is annihilated while a photon is generated. In our experiments, the optical cavity has a decay rate $\kappa$ that is much faster than $G$, and its photons are sent to a detector. Instead of coherent dynamics between the optical field and motion, detection leads to a quantum operation with corresponding jump operators proportional to $\bopd$ and $\bop$ for the two Hamiltonians, respectively. Therefore, the detection of an individual photon at the cavity resonance frequency heralds the addition or subtraction of a phonon to the mechanical mode characterized by the operation of the respective jump operator. The state of the mechanical oscillator will be thermal before the record of detections is taken into account. This reflects our lack of knowledge about the motional state and that the mechanical oscillator is in equilibrium with its environment. Starting with the thermal density matrix $\rhothermal$, detection of a photon corresponds to updating the state with the jump operator corresponding to the correct detuning. For the blue-detuned case, we use the jump operator proporational to $\bopd$ to obtain the phonon-added state: $\rhothermal \rightarrow \bopd \rhothermal \bop$. In the red-detuned case, we have phonon-subtraction represented by: $\rhothermal \rightarrow \bop \rhothermal \bopd$.

Our experimental setup is shown in simplified form in Fig.\ref{fig:setup}b. First, we send light from a laser into an optomechanical crystal cavity, either red-detuned or blue-detuned from the cavity resonance by the mechanical frequency. The fiber to chip coupling efficiency is $\etaf \approx\SI{76}{\percent}$. We split off some of the light before interacting with the device, to use as a local oscillator. A delay is applied using about 100 meters of fiber to approximately compensate for the signal path. A continuous wave, frequency upshifted ($\SI{40}{\mega\hertz}$), probe tone is generated using an acousto-optic-modulator. The reflected light from the cavity is then split into two paths. One path, for single photon counting, contains two cascaded high finesse fiber fabry-perot cavity filters with free spectral range of $\SI{15}{\giga\hertz}$ and finesse of 300 (Micron Optics FFP-I) to suppress the pump light and pass through only the photons due to scattering from the mechanical resonator. Another path, for heterodyne detection, contains a balanced heterodyne receiver (Thorlabs $\SI{75}{\mega\hertz}$ PDB425C-AC) whose output is sent to a $\SI{12}{\bit}$ $\SI{500}{\mega\sample\per\second}$ digitizer (Alazartech ATS 9350). The local oscillator and signal are combined on a variable optical coupler before the balanced detector. The DC output level of the two photo-diodes that comprise the receiver is monitored, and the coupler splitting ratio is adjusted until the voltages are approximately equal. We note that the phase of our local oscillator is left unlocked. This simplifies the experiment but means that we extract no information about the phase of the system, causing our inferred states to have rotationally symmetric quasi-probability distributions. Our justification is that the initial state is a rotationally symmetric thermal state, and that the photon addition and subtraction processes occur at random times and have no associated phase. 

In the photon counting path, we send the light reflected from the room temperature device to a superconducting nanowire single photon detector (SNSPD) which resides on the still plate of a dilution refrigerator (Bluefors) at $\approx \SI{1}{\kelvin}$. The SNSPD (Photon Spot) has an effective quantum efficiency of about $\SI{70}{\percent}$ (including all fiber losses leading into the fridge) and dark counts on the order of $\SI{30}{\hertz}$ when biased with $\SI{10}{\micro\ampere}$.

Our optomechanical crystal device, used to read out and control mechanical motion, is similar to those presented in previous work~\cite{Patel2017a}. The optical cavity mode has a center wavelength of $\lambda_{\textrm{c}} = \SI{1551}{\nm}$, total decay rate $\kappa/2\pi = \SI{822}{\mega\hertz}$, and external coupling rate of $\kappae/2\pi = \SI{190}{\mega\hertz}$. The probability that a cavity photon leaks into the detected waveguide channel is given by the cavity efficiency, $\etac = \kappae/\kappa \approx \SI{23}{\percent}$. The mechanical mode frequency is $\omegam/2\pi \approx \SI{3.96}{\giga\hertz}$. From a sweep of laser power, we determine the intrinsic, backaction-free, mechanical linewidth of $\gammai/2\pi = \SI{2.06\pm0.01}{\mega\hertz}$ and single-photon optomechanical coupling rate $g_{0}/2\pi = \SI{1.01\pm0.02}{\mega\hertz}$. The measured value of $g_{0}$ deviates by about $\SI{5}{\percent}$ from the simulated $g_{0}/2\pi = \SI{961}{\kilo\hertz}$, a difference which we can attribute to systematic errors in power calibration as well as uncertainties in the material's photoelastic parameters. In terms of the system parameters above, the single-photon generation rate per phonon in the mechanical resonator, is proportional to the optomechanical measurement rate: $\gammaOM = 4 g^{2}_{0} n_{\textrm{cav}}/\kappa$, where $n_{\textrm{cav}}$ is the number of optical intracavity photons (on order $10^{2}$ in this experiment).

\begin{figure}
\includegraphics{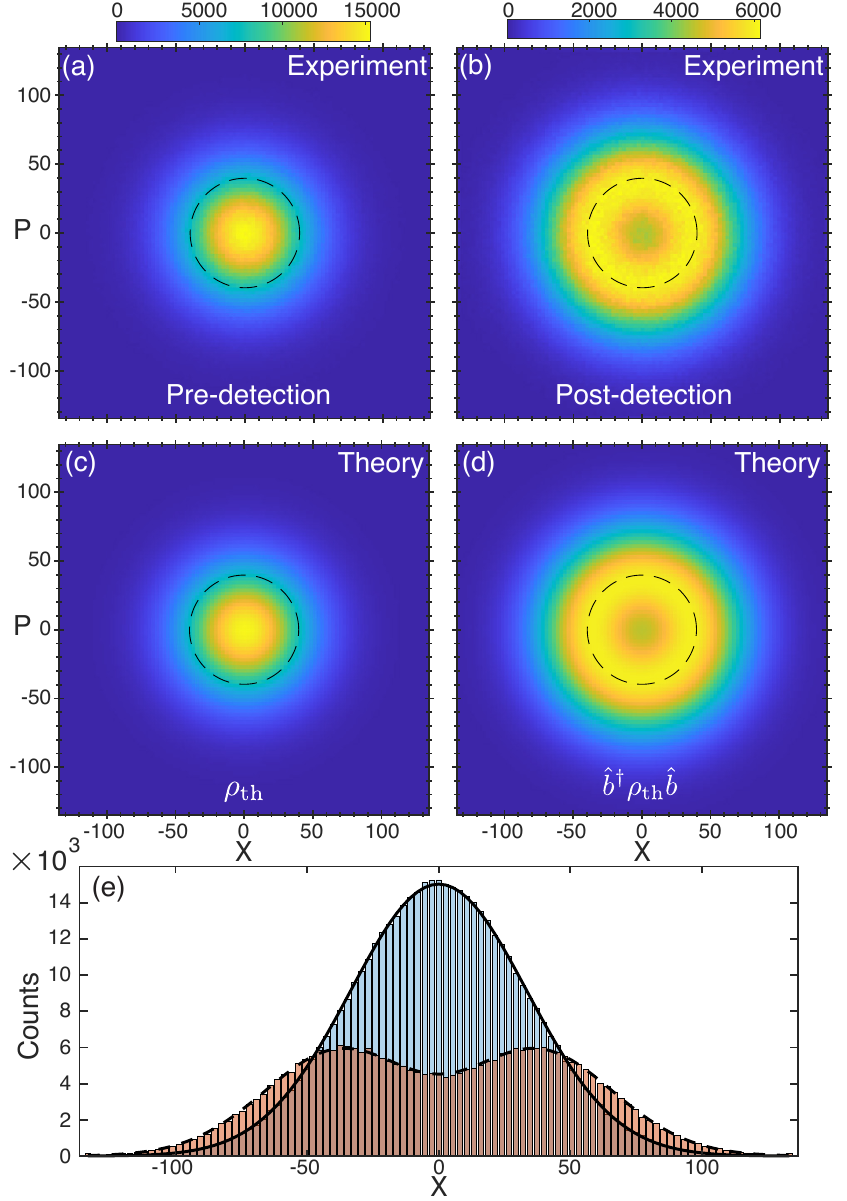}
\caption{\label{fig:histograms} \textbf{Measurement of quadrature histograms for thermal and phonon-added thermal state} (a) Experimental result showing a histogram of $1.5\times10^{7}$ quadrature samples. The data are binned into 101 bins along the $X$ and $P$ axes. Color bars show counts per bin. The statistics are Gaussian, corresponding to a mechanical thermal state. (b) Post-selected results, triggered by single-photon clicks. The observed statistical distribution now corresponds to a phonon-added state and is non-Gaussian. (c,d) Corresponding theory plots to (a,b), obtained by fitting the analytically determined Husimi Q functions via two parameters. The plots shown are the binned theory Q functions for both the thermal and postselected case. The dotted black circles, shown for reference in the plots, have radii $r = \sqrt{\bar{n}_{\textrm{th}}}$, the characteristic length scale for thermal fluctuations. $\bar{n}_{\textrm{th}} \approx 1580$ phonons, corresponding to the mean occupancy of a $\SI{3.96}{\GHz}$ oscillator at room temperature. (e) Blue (red) bars correspond to a linecut of the 2D experiment histograms at $P = 0$ for the thermal (post-selected) datasets. The theory fits for the thermal (post-selected) results are shown in solid (dotted) black lines.}
\end{figure}

The first experiment we perform is phonon-addition. We tune the laser on the blue side of the optical cavity resonance. The experimental data is collected by triggering the digitizer on a single photon click, collecting $\approx\SI{2}{\micro\second}$ of heterodyne data, and estimating the in-phase and in-quadrature components of the down-converted mechanical signal. The resulting complex voltage samples, $v = \sqrt{G}(X + iP)$ comprise the dataset with which we perform tomography. Here, $G$ is the overall detection gain and $X$ and $P$ are the in-phase and in-quadrature components respectively. We collect the quadrature samples in two interleaved phases, one in which data collection is triggered by a single photon click, the other in which the clicks are ignored. When the measurement is triggered by a single photon, a phonon-added thermal state is heralded. The histogram of quadrature samples, obtained by binning the raw data for a mechanical thermal state, is shown in Fig.~\ref{fig:histograms}a. These data were binned into 101 bins in both the $X$ and $P$ directions. The tomography of the thermal state shows a Gaussian distribution of quadrature amplitudes (Fig.~\ref{fig:histograms}a). By contrast,  we observe a clear non-Gaussian rotationally symmetric distribution in the phonon-added thermal state in Fig.~\ref{fig:histograms}b.

The Husimi Q function for the post-selected phonon-added thermal state has the form:
\begin{equation}
\label{eq:formQ}
    Q_{\textrm{post}}(\alpha) \propto \bra{\alpha}\bopd \hat{\rho}_\textrm{th} \bop\ket{\alpha} \propto |\alpha|^{2}e^{-|\alpha|^{2}/(\bar{n}_{\textrm{th}}+1)},
\end{equation}
where $\bar{n}_{\textrm{th}}$ is the mean thermal phonon occupancy and $\alpha = X + iP$. $Q_{\textrm{post}}(\alpha)$ describes the measurement statistics in phase space, in the absence of technical noise. Since the $Q$ function is a probability distribution, the effect of added Gaussian noise can be represented by its convolution with a Gaussian distribution $\mathcal{N}$, with zero mean and a variance of $n_{\textrm{added}}$:
\begin{equation}
\label{eq:convQ}
    Q(\alpha) =  \Big(Q_{\textrm{post}} \ast \mathcal{N}\Big) (\alpha).
\end{equation}
(See Appendix for an analytic expression for the measured $Q(\alpha)$ which depends only on $\bar{n}_{\text{th}}$ and $n_{\textrm{added}}$). To relate to experimental observation, an additional scaling parameter $G$ is needed. This distribution is binned, and fit to the data via a maximum-likelihood method. Since the bath temperature is known ($T\approx \SI{300}{\kelvin}$), we fix $\bar{n}_{\textrm{th}} = 1578$ in Eq.~\ref{eq:formQ} and fit Eq.~\ref{eq:convQ} via two free parameters: detection gain $G$, and number of added noise phonons $n_{\textrm{added}}$. The parameter $G$ is used to re-scale the data as $v=\sqrt{G}\alpha$. The result of the fit with $(G,n_{\textrm{added}})=(8.1\times10^{-3}~\text{{V}}^{2},670)$ is shown in Fig.~\ref{fig:histograms}c,d, in good agreement with the experimental results of Fig.~\ref{fig:histograms}a,b for both pre- and post-selected distributions. We note that the thermal datasets are interleaved with the post-selection data sets to mitigate the effects of drift. Furthermore, no additional fitting is performed on the thermal datasets, confirming the validity of our gain and added noise estimates. Figure~\ref{fig:histograms}e shows a line-cut of the 2D histograms with fit results for the thermal (solid line) and post-selected (dotted line) distributions respectively. 

\begin{figure}
\includegraphics{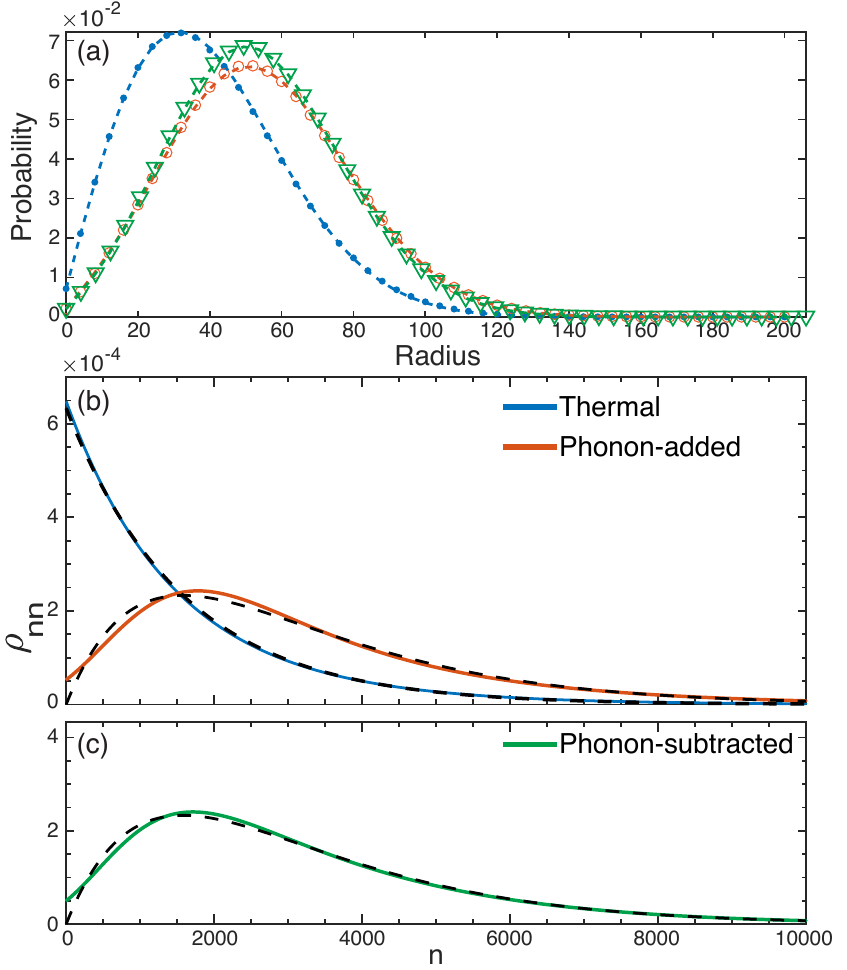}
\caption{\label{fig:densitymatrix}\textbf{Reconstructed density matrix elements} (a) Radial histogram of measurement results for the thermal state (filled blue points), the phonon-added thermal state (open red circles) and the phonon-subtracted thermal state (green triangles). The respective dotted lines show the theory fits. (b) Diagonal density matrix elements reconstructed from experimental data. Estimates for gain and added noise are provided by the fits to the quadrature histogram data. The black dotted lines show the result of theory for the thermal and phonon-added states with thermal bath occupancy set to $\bar{n}_{\textrm{th}} = 1578$ phonons ($\approx \SI{300}{\kelvin}$ mode temperature). (c) Experimentally reconstructed density matrix elements for a phonon-subtracted thermal state.}
\end{figure}

Although we have considered two dimensional histograms of our data so far, a more compact representation can be achieved by noting the radial symmetry of the generated states. The results of radial binning for the thermal and phonon-added states are shown in (Fig.~\ref{fig:densitymatrix}a). Here, blue (red) points show thermal (phonon-added) results respectively, while the dotted lines show the theoretical fits. In addition, we perform phonon-subtraction by tuning our pump laser to the red side of the optical cavity. The radial statistics for the resulting noise distribution is shown by the open green triangles. As expected from theory, the result is very similar to the case of phonon-addition. 

To further analyze the data, we reconstruct the density matrix describing the mechanical system. From the radial histograms, we estimate the density matrix using an iterative maximum-likelihood method method~\cite{lvovsky_iterative_2004,eichler2012characterizing}. We simplify the optimization by restricting ourselves to diagonal density matrices, as justified by the radial symmetry of the phase space distribution (see Appendix for a complete discussion of these techniques). The results are shown in Fig.~\ref{fig:densitymatrix}b. From this reconstruction, we obtain a mean phonon number of $\bar{n}_{\textrm{th}} = 1597.6\pm0.6$ and $\bar{n}_{\textrm{post}} = 3158.1\pm0.9$ for the thermal and post-selected states respectively. The probability of the vacuum component is reduced markedly post-addition of a phonon, going from $p_\textrm{vac,th} = (6.49\pm0.01)\times10^{-4} $ to $p_\textrm{vac,post} = (5.25\pm0.06)\times10^{-5} $. In all quantities estimated from the state reconstruction, the quoted errors reflect statistical uncertainty obtained using a bootstrapping method of the entire dataset. This method works by re-sampling the entire dataset of $1.5\times10^{7}$ samples with replacement 50 times, and reconstructing the density matrix for each trial. Lastly, we perform the reconstruction of the phonon-subtracted state, shown in (Fig.~\ref{fig:densitymatrix}c).

Next, we use our reconstruction results to investigate how the expected number of phonons changes after post-selection. We compute the ratio of the two mean phonon numbers as $\bar{n}_{\textrm{post}}/\bar{n}_{\textrm{th}} = 1.977 \pm 0.001$. This is in agreement with theory, in which the mean phonon number for both the phonon-added and -subtracted thermal states should approximately be twice that of the original thermal state. This counter-intuitive result that adding \emph{or} {subtracting} a phonon \emph{doubles} the mean number of phonons in a resonator, $\bar{n}_{\textrm{post}} \approx 2\bar{n}_{\textrm{th}}$, is best understood by considering the information gained about the mechanical system from the optical single photon measurement. Before the measurement of a photon occurs, the \emph{a-priori} probability distribution over each phonon energy level, $n$, is given by the familiar exponentially decaying Boltzmann factor: $P(n) \propto e^{-\beta n}$ where $\beta$ is the inverse temperature (Fig.~\ref{fig:densitymatrix}b, blue curve). Once a click has occurred however, the observer gains information about the state, and we must update these probabilities via Bayes' rule. Letting the number of resonant cavity photons in a small time interval be $N$, we have the following update rule: $P(n|N=1) \propto P(N=1|n)P(n)$, a rescaling of the \emph{a-priori} distribution. Now, the probability of a photon scattering event itself depends on the phonon number: $P(N=1|n) \propto n$. Thus we see that the \emph{a-posteriori} probability is the prior distribution re-scaled by $n$, leading to the suppression of probability for small phonon numbers (Fig.~\ref{fig:densitymatrix}b). This causes the average phonon number to be increased. 

One can verify this intuitive argument by direct calculation of the phonon-added state. Writing the thermal state as a sum (for $n \geq 0$) using the prior, thermal, probability distribution: $\rhothermal = \sum_{n} P(n) \ketbra{n}{n}$, we calculate the post-selected phonon-added state as $\hat{\rho}_{\textrm{post}} = (\bopd \rhothermal \bop)/ \Tr(\bop\bopd\rhothermal)$. Recalling that $\bopd \ket{n} = \sqrt{n+1} \ket{n}$ and simplifying, we get: $\hat{\rho}_{\textrm{post}} = (1/\nbarthermal) \sum_{n} n P(n)\ketbra{n}{n}$. Notice that the priors have been updated: $P(n) \rightarrow nP(n)/\nbarthermal$. This analysis, while illustrated for phonon-addition, applies as well to phonon-subtraction in the large thermal occupation limit (see Eq.~\ref{eq:phonsub}). 

As noted above we observe a near, though inexact, doubling in mean phonon number after post-selection. To understand this discrepancy with theory, we independently measure the dark count rate in our measurement. Dark counts introduce a loss in fidelity of the heralded state. More precisely, the heralding fidelity is defined by the quantity $\chi = \Gamma_{\textrm{sig}}/(\Gamma_{\textrm{sig}} + \Gamma_{\textrm{dark}})$. Here, $\Gamma_{\textrm{sig}}$ denotes the count rate of thermal signal phonons, while $\Gamma_{\textrm{dark}}$ denotes the total dark count rate, caused by a sum of intrinsic SNSPD dark counts and pump feedthrough from imperfect pump rejection filtering. We have independently estimated these quantities, using the measured photon count rate on-resonance, and off-resonance. We measure $\Gamma_{\textrm{sig}} \approx \SI{278\pm5}{\kilo\hertz}$ and $\Gamma_{\textrm{dark}} \approx \SI{5.2\pm0.1}{\kilo\hertz}$. From this we estimate the fidelity, $\chi \approx 0.98\pm0.02$. Written in terms of fidelity, the theoretically expected ratio of mean occupancy is: $\bar{n}_{\textrm{post}}/\bar{n}_{\textrm{th}} \approx 1 + \chi$. The measured fidelity $\chi$ thus closely explains the observed ratio $\bar{n}_{\textrm{post}}/\bar{n}_{\textrm{th}}$. 

\begin{figure}
\includegraphics{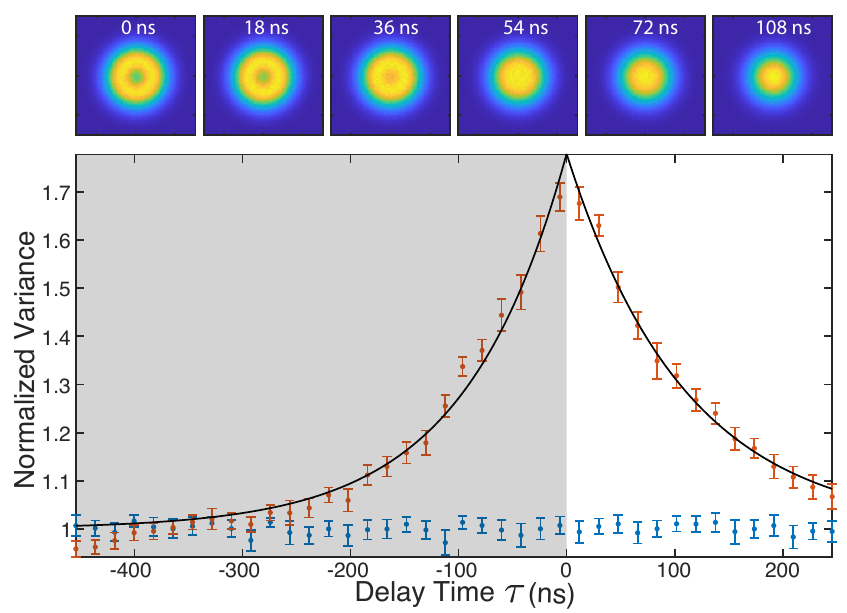}
\caption{\label{fig:correlations}\textbf{Time-evolution of phonon-added state} The top insets from left to right show how the raw quadrature histograms change as the delay time $\tau$ is swept. The noise variance, normalized to the variance of the thermal state, is plotted below as a function of $\tau$. Red (blue) points denote the results for post-selected (thermal) distributions respectively. The temporal mode-matching between the single photon pulse and the heterodyne signal is well defined for $\tau \geq 0$. As such we shade out the negative delay region, in which the matched filter of the detection chain only partially overlaps with the decaying mechanical signal.}
\end{figure}

Finally, we map out the time-evolution phonon-added mechanical thermal state. We sweep the delay time $\tau$, which controls the temporal mode matching between the signal at the heterodyne detector and the single photon counter (Fig.~\ref{fig:setup}b). As a function of delay, we compute the total variance of the quadrature histograms. An exponential decay is observed, with a decay time on the order of the mechanical lifetime ($\tau \approx \SI{100}{\nano\second}$). We observe that the noise distribution near zero delay is distinctly non-Gaussian, but tends towards a Gaussian thermal distribution at large delay. The results, normalized to the variance of the mechanical thermal noise Gaussian, are shown in Fig.~\ref{fig:correlations}.  

In this work we have experimentally demonstrated phonon addition and subtraction followed by state tomography in an optomechanical system. These capabilities open up several directions for future studies. First of all, the single-phonon-added thermal states demonstrated here have theoretically been shown to be non-classical at all temperatures. This is ensured by the negativity of their Glauber-Sudarshan P functions~\cite{agarwal_nonclassical_1992,Kiesel,Zavatta}. Although this negativity is difficult to detect experimentally at room temperature, we hope that our work will motivate further studies into whether weakly non-classical, but non-Gaussian, states may prove useful as a quantum resource~\cite{Zhuang2018,Genoni2013}. Finally, we point out that our experiment demonstrates the ability to add a node to the Q function of a single-mode oscillator. Since it is known that any pure state whose Q function contains nodes is non-classical~\cite{chabaud2020stellar}, one could prepare non-classical states beyond the phase insensitive ones we have reconstructed. In particular, we expect with near-term improvements, including the implementation of phase-sensitive detection, our technique will allow the preparation of phonon-added coherent states of motion~\cite{Li2018,agarwal_nonclassical_1991}. Likewise, phonon-subtraction, performed on a squeezed mechanical steady state, gives a route to cat state generation in the mechanical domain~\cite{Kronwald2013,ourjoumtsev2006generating}.

During the preparation of this manuscript we became aware of related work demonstrating variance doubling in phonon-added and subtracted mechanical thermal states \cite{enzian2021single}.

\begin{acknowledgments}
This work was funded by the U.S. government through the Department of Energy through Grant No. DE-SC0019174, and the U.S. Army Research Office (ARO)/Laboratory for Physical Sciences (LPS) Cross-Quantum Systems Science \& Technology (CQTS) program (Grant No. W911NF-18-1-0103). The authors wish to thank NTT Research for their financial and technical support. We acknowledge the David and Lucille Packard Fellowship, and the Stanford University Terman Fellowship. We thank Pieter-Jan C. Stas, Alex Wollack, Hubert Stokowski and Marek Pechal for experimental support. Device fabrication was performed at the Stanford Nano Shared Facilities (SNSF) and the Stanford Nanofabrication Facility (SNF). The SNSF is supported by the National Science Foundation under Grant No. ECCS-2026822. RNP was partly supported by the National Science Foundation Graduate Research Fellowship under Grant No. DGE-1656518.
\end{acknowledgments}

\section*{Appendix}
\appendix
\section{Phonon-added and subtracted states}

In this section we state results for the density matrix elements of phonon added and subtracted states. We also give analytic expressions of the Husimi Q function, fundamental to analyzing the output of heterodyne tomography experiments, both for the noiseless (zero technical noise) and added-noise cases. 

\subsection{Density Matrix}
In the formulas below, we use the shorthand notation $\beta = 1/kT$ and $\hbar\omega = 1$. 
The mechanical thermal state is given by: 
\begin{equation}
\label{rhothermal}
    \rhothermal = \frac{1}{\bar{n} + 1} \sum_{n=0}^{\infty} \exp(-\beta n) \ketbra{n}{n}
\end{equation}
where $\Tr[\hat{n}\rhothermal] = \bar{n} = 1/(\exp(\beta) - 1)$ is the mean phonon occupancy. Following post-selection, the phonon-added state is: 
\begin{equation}
\label{added}
\hat{\rho}_{\textrm{a}} = \frac{\bopd\rhothermal\bop}{\Tr[\bop \bopd \rhothermal]} = \frac{1}{\bar{n}(\bar{n} + 1)} \sum_{n=0}^{\infty} n \exp(-\beta n) \ketbra{n}{n}.
\end{equation}
In this state, the mean phonon number is $\Tr[\hat{n}\hat{\rho}_{\textrm{a}}] = 2\bar{n} +1$. The phonon-subtracted state is: 
\begin{equation}
   \hat{\rho}_{\textrm{s}} = \frac{\bop\rhothermal\bopd}{\Tr[\bopd \bop \rhothermal]} = \frac{1}{(\bar{n} + 1)^{2}} \sum_{n=0}^{\infty} (n+1) \exp(-\beta n) \ketbra{n}{n} 
\end{equation}
or: 
\begin{equation}
\label{eq:phonsub}
    \hat{\rho}_{\textrm{s}} = \frac{1}{\bar{n} + 1} \left( \rhothermal + \bar{n} \hat{\rho}_{\textrm{a}} \right)
\end{equation}
Note that for $\bar{n} \gg 1$, as in this experiment, $\hat{\rho}_{s} \approx \hat{\rho}_{a}$

\subsection{Husimi Q functions for phonon-added and subtracted states}
The Husimi Q function is defined as: 
\begin{equation}
    Q(\alpha) = \frac{1}{\pi} \bra{\alpha} \hat{\rho} \ket{\alpha}.
\end{equation}
Substituting the three states of interest into this equation give the following results. 

For a thermal state with mean phonon number $\bar{n}$ the Q function is given by: 
\begin{equation}
    Q_{\textrm{th}}(\alpha) = \frac{1}{\pi(\bar{n} + 1)}\exp (\frac{-|\alpha|^{2}}{\bar{n}+1}). 
\end{equation}

The phonon-added state Q function is: 
\begin{equation}
    Q_{\textrm{a}}(\alpha) = \frac{1}{\pi(\bar{n} + 1)^{2}}|\alpha|^{2}\exp (\frac{-|\alpha|^{2}}{\bar{n}+1}).
\end{equation}
Note in particular that this distribution is no longer Gaussian, and goes to 0 at $\alpha = 0$ for all temperature (all values of $\bar{n}$).

The phonon subtracted Q function can be expressed as a weighted sum of the previous two functions. Stated explicitly: 
\begin{equation}
\label{Qsub}
    Q_{s}(\alpha) = \frac{1}{\pi(\bar{n} + 1)^{2}} \left(1 + \bar{n}|\alpha|^{2} \right)\exp (\frac{-|\alpha|^{2}}{\bar{n}+1}).
\end{equation}
This equation implies that the non-Gaussian character of the phonon-subtracted state increases with increasing temperature.

\subsection{Husimi Q functions in presence of technical noise}
The detected Q function for the post-selected state, in the presence of technical noise is: 
\begin{equation}
\label{eq:2Dpostselconv}
    Q(r) = \frac{1}{4\pi} \left( \frac{1}{\sigma_{2}\sigma_{1}^2} \right)^{2} \left( 2 \Tilde{\sigma}^{4} +  \frac{\Tilde{\sigma}^{6}}{\sigma^{4}_{2}} r^2\right)  \exp (\frac{-r^{2}}{2(\sigma^{2}_{1} + \sigma^{2}_{2})}),
\end{equation}
where $\sigma^{2}_{1} = (\bar{n} + 1)/2$ is the variance in the noise of the state, and $\sigma^{2}_{2} \approx n_{\textrm{added}}/2$ is the variance of the added technical noise. 

\begin{equation}
    \Tilde{\sigma}^{2} = \frac{\sigma^{2}_{1} \sigma^{2}_{2}}{\sigma^{2}_{1} + \sigma^{2}_{2}}
\end{equation}

\section{Derivation of Time-Independent Fields after Filtering} 

In this section, we present an input-output theory analysis of the detected noise, and show the dependence of detected noise on the matched filter bandwidth and delay time. We show the exponential decay of the signal variance with time, which we have detected in experiment.

\subsection{Output fields from input-output theory}
The output cavity field is sampled continuously in this experiment. The output of the detector is a time independent quantity, obtained by integrating the field with a filter function. Our treatment of this problem is similar to \cite{eichler_experimental_2011}, although we work in a regime in which thermal noise from the mechanical mode plays a significant role. In the following we derive expressions for the integrated output cavity field: 
\begin{equation}
    \label{timeindep}
    \Aout = \int f(t) \Op{a}{\textrm{out}} (t) dt
\end{equation}
And the resulting measured photocurrent: 
\begin{equation}
    \hat{I} = \alpha_{\textrm{LO}} (\Op{A}{\textrm{out}} + \Opd{A}{\textrm{out}})
\end{equation}
where $\alpha_{\textrm{LO}}$ denotes the local oscillator strength (whose phase is random in our experiment, but which we take to be real for simplicity), and the filter function is given by: 

\begin{equation}
    f(t) = \Theta (t - \tau) \sqrt{\beta}  \exp({-\frac{\beta}{2} (t - \tau)}) . 
\end{equation}

In the above equation $\beta$ denotes the matched filter energy decay rate, and the function $\Theta(t-\tau)$ is the Heaviside step function starting at time $\tau > 0$. 

The filter satisfies the normalization condition: 

\begin{equation}
    \int_{-\infty}^{\infty} \norm{f(t)}^{2} dt = 1.
\end{equation}

To proceed, we write the Heisenberg-Langevin equations describing the dynamics of the system operators in a frame rotating at the mechanical frequency:
\begin{equation}
\label{HLoptics}
    \ddt {\aop} (t) = -\frac{\kappa}{2} \aop(t)- iG \bop(t) - \sqrt{\kappae}\ain(t) - \sqrt{\kappai}\aini(t),
\end{equation}
\begin{equation}
\label{HLmech}
    \ddt {\bop} (t) = -\frac{\gamma}{2} \bop(t) - \sqrt{\gammai} \bin(t) + \hat{F}_{BA}(t)
\end{equation}
along with the input-output boundary condition: 
\begin{equation}
\label{HLoutput}
    \aout(t) = \ain(t) + \sqrt{\kappae}\aop(t). 
\end{equation}
Note that in Eq.~\ref{HLoptics} the final term arises from vacuum noise due to undetected channels \cite{Safavi-Naeini2013f}. In Eq.~\ref{HLmech} we have taken into account the effect of the optical mode into the mechanical damping rate by writing $\gamma = \gammai \pm \gammaOM$,
where $\gammaOM = \frac{4G^{2}}{\kappa}$
is the optomechanical measurement rate, and the minus sign is chosen for blue laser-cavity detuning.

Assuming $\kappa \gg G,\gamma,\beta$ throughout, we set the left hand side of Eq.~\ref{HLoptics} to 0 and substitute $\bop(t)$ with the formal solution of Eq.~\ref{HLmech}: 
\begin{equation}
    \bop(t) = \bop(0) \exp(-\frac{\gamma}{2} t) - \sqrt{\gammai} \int_{0}^{t} \exp(-\frac{\gamma}{2} (t-t')) \bin(t')dt'
\end{equation}
to obtain an equation for $\aop(t)$ and $\aout(t)$:
\begin{widetext}
\begin{equation}
\aop(t) = -\frac{2iG}{\kappa} \left[ \bop(0) \exp(-\frac{\gamma}{2} t) - \sqrt{\gammai} \int_{0}^{t} \exp(-\frac{\gamma}{2} (t-t')) \bin(t')dt'  \right] - \frac{2\sqrt{\kappae}}{\kappa} \ain(t) - \frac{2\sqrt{\kappai }}{\kappa}\aini(t)
\end{equation}
\begin{equation}
\label{aoutfull}
    \aout(t) =-i\sqrt{\etac \gammaOM} \left[ \bop(0) \exp(-\frac{\gamma}{2} t) - \sqrt{\gammai} \int_{0}^{t} \exp(-\frac{\gamma}{2} (t-t')) \bin(t')dt'  \right] + (1 - 2\etac) \ain(t) - \frac{2\sqrt{\kappai \kappae}}{\kappa}\aini(t)
\end{equation}
\end{widetext}
where in Eq.~\ref{aoutfull} we substituted the definition of cavity efficiency $\etac = \kappae/\kappa$ and $\gammaOM$. We have ignored here the contribution due to the optical read-out field on the mechanical motion $\hat{F}_{BA}(t)$, as this back-action noise is much smaller than the noise we will be measuring~\cite{Safavi-Naeini2013f,Khalili2012}. The first two terms of this equation represent contributions from the decaying mechanical mode and the thermal noise respectively. The last two terms represent contributions from optical vacuum noise arising from both driven and undetected channels. 

To obtain the time independent field via Eq.~\ref{timeindep} we evaluate: 

\begin{widetext}
\begin{equation}
\begin{split}
        \Aout = -i\sqrt{\beta \etac \gammaOM} \exp(\beta \tau/2) \Bigg[ \Bigg. \bop(0) \int_{0}^{\infty}\Theta(t - \tau) \exp(-\frac{\gamma + \beta}{2} t) dt \\ - \sqrt{\gammai}\int_{0}^{\infty} \int_{0}^{t} \Theta(t - \tau)\exp(-\frac{\gamma + \beta}{2} t) \exp(\frac{\gamma}{2}t') \bin(t')dt'dt  \Bigg. \Bigg] \\
        + (1 - 2\etac) \exp(\beta \tau/2) \int_{0}^{\infty}\Theta(t - \tau)\exp(-\frac{\beta}{2}t)\ain(t) dt \\
        - \frac{2\sqrt{\kappai \kappae}}{\kappa} \exp(\beta \tau/2) \int_{0}^{\infty}\Theta(t - \tau)\exp(-\frac{\beta}{2}t)\aini(t) dt
\end{split}
\end{equation}

which reduces to: 

\begin{equation}
\label{aouttimeindep}
\begin{split}
        \Aout =-\frac{2i\sqrt{\beta \etac \gammaOM}}{\gamma + \beta} \Bigg[ \Bigg. \bop(0)\exp(-\frac{\gamma}{2}\tau) - \sqrt{\gammai} \exp(\beta\tau/2)\int_{\tau}^{\infty} \exp(-\frac{\beta}{2} t)\bin(t)dt - \sqrt{\gammai} \exp(-\gamma\tau/2)\int_{0}^{\tau} \exp(\frac{\gamma}{2} t)\bin(t)dt \Bigg. \Bigg] \\
        + \sqrt{\beta}(1 - 2\etac) \exp(\beta \tau/2) \int_{\tau}^{\infty}\exp(-\frac{\beta}{2}t)\ain(t) dt \\
        - \frac{2\sqrt{\beta\kappai\kappae}}{\kappa} \exp(\beta \tau/2) \int_{\tau}^{\infty}\exp(-\frac{\beta}{2}t)\aini(t) dt.
\end{split}
\end{equation}
\end{widetext}

This equation allows us to compute the fluctuations of the photocurrent as a function of matched filter linewidth and start time: 
\begin{equation}
    \label{currentsquared}
    \langle \Opd{I}{} \Op{I}{} \rangle (\tau,\beta) = \alphaLO^{2}\langle(\Aout + \Aout^{\dagger})^{2}\rangle.
\end{equation}
Neglecting cross-correlations that appear in Eq.~\ref{currentsquared}, we evaluate all expectation values in angular brackets given a state $\rho = \hat{\rho}_{\textrm{a}} \bigotimes \hat{\rho}_{\textrm{thermal}}$, where $\hat{\rho}_{\textrm{a}}$ is the post-selected state after heralding, given by Eq.~\ref{added}, and where all input operators act on the $\hat{\rho}_{\textrm{thermal}}$ state. In our derivation we note that the expectation value of the system operators in the post-selected state is: $\langle \bopd(0) \bop(0)\rangle \approx 2\bar{n}\frac{\gammai}{\gamma}$ .(It differs from $2\bar{n}$ due to backaction from the laser drive). The final result for $\tau \geq 0$ and $\bar{n} \gg 1$ is:
\begin{equation}
\begin{split}
     \langle \Opd{I}{} \Op{I}{} \rangle (\tau,\beta) = \frac{8\alphaLO^{2}\etac\gammaOM}{(\gamma + \beta)^{2}} \bar{n}\gammai \Big[\frac{\beta}{\gamma}(1+ e^{-\gamma\tau}) + 1\Big] +\alphaLO^{2}.
\end{split}
\end{equation}
The first set of terms in the brackets represent contributions from the state and the input thermal noise. The last term denotes a constant noise floor due to the optical vacuum noise (shot noise). Notice that the fluctuations at 0 delay roughly double compared to their large-delay value in the limit where $\beta \gg \gamma$. In our experiment, the matched filter bandwidth is limited by preceding filters. If $\beta$ is made too large, then the shot-noise term (and other added technical noise contributions) begin to dominate the measurement signal.

\section{Detector and Device Characterization}

In this section we present details related to our heterodyne down-conversion scheme, and device characterization of the mechanical mode in our experiment. 

\begin{figure}
    \includegraphics{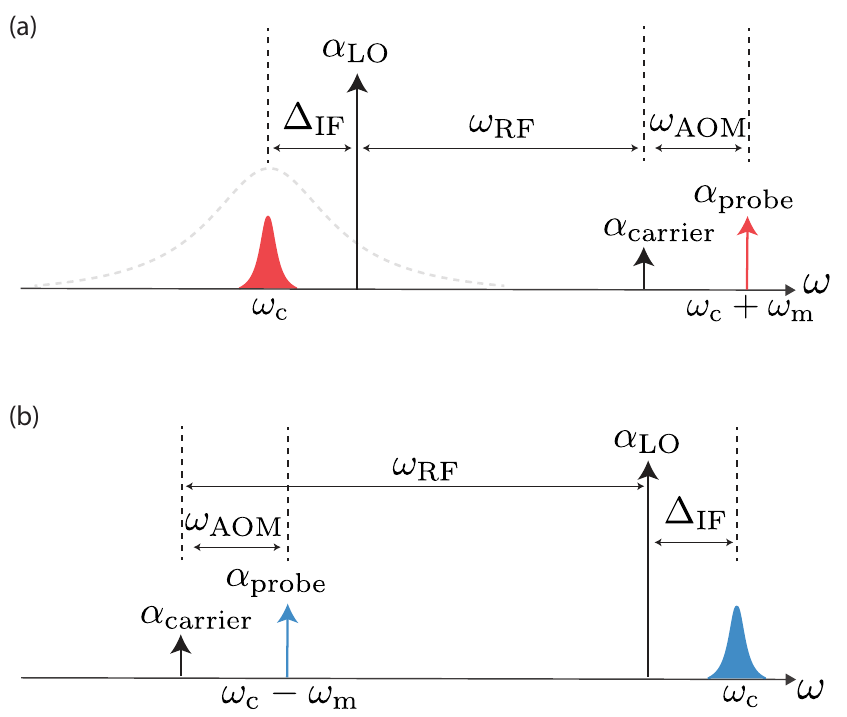}
    \caption{Heterodyne scheme: generating single-sideband local oscillator for detecting mechanical motion. (a) Schematic showing the down-conversion scheme with probe (vertical red arrow) tuned on the blue side of the optical cavity (grey dotted curve). The local oscillator tone,$\alphaLO$, beats against the mechanical signal generating a signal at $\Delta_{\textrm{IF}}$. (b) Schematic for the case where the probe tone is red-detuned from cavity resonance.}
    \label{fig:SSB}
\end{figure}

We perform down-conversion of the mechanical signal by shifting the frequency of our laser using single-sideband suppressed-carrier modulation. Such a modulation scheme avoids added shot-noise from sidebands that do not contribute to signal gain. We implement the single-sideband modulation using a quadrature phase shift keying (QPSK) modulator (Optilab QPSK-OM-23). Figure~\ref{fig:SSB} describes the relevant tones in frequency domain. In this diagram, the vertical arrows denote the principal laser frequencies used in the experiment, where $\alphaLO$ denotes the local oscillator tone generated from modulating the carrier, $\alpha_{\textrm{carrier}}$ denotes the carrier tone prior to any modulation, and $\alpha_\textrm{probe}$ denotes the probe which is sent to the optomechanical crystal. The probe is up-shifted from the carrier by $\omega_{\textrm{AOM}}/2\pi \approx \SI{40}{\mega\hertz}$ using an acousto-optic modulator. In Fig.~\ref{fig:SSB}a, we show the schematic for phonon-addition, where the probe is blue-detuned from the cavity frequency $\omegac$ by a mechanical frequency $\omegam$. In order to down-convert the mechanical signal to a chosen intermediate frequency $\Delta_\textrm{IF}$, we generate a down-shifted RF drive at the frequency: 

\begin{equation}
    \omega_{\textrm{RF}} = \omegam - \Delta_{\textrm{IF}} - \omega_{\textrm{AOM}}
\end{equation}

Similarly, for phonon-subtraction and red-side driving of the optical cavity, inspection of Fig.~\ref{fig:SSB}b gives the required RF frequency for upconversion:

\begin{equation}
         \omega_{\textrm{RF}} = \omegam - \Delta_{\textrm{IF}} + \omega_{\textrm{AOM}}.
\end{equation}

\begin{figure}
    \includegraphics{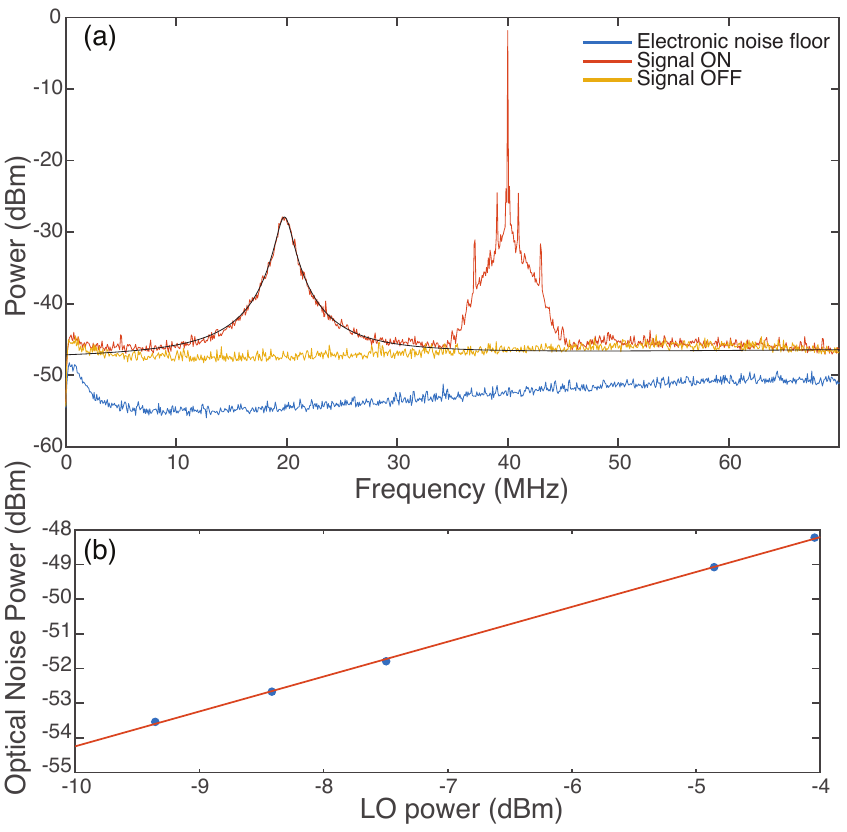}
    \caption{(a) Noise power spectra obtained at the output of the balanced photodetector. Total power in a $\SI{25}{\kilo\hertz}$ bandwidth is shown vs. frequency. The blue curve shows the noise level when the optical local oscillator is off. With the local oscillator on, but the signal port blocked, we observe an increase in the noise floor due to shot-noise (yellow curve). The red curve shows the heterodyned signal. Mechanical thermal noise from the device is visible as a peak at $\SI{20}{\mega\hertz}$, where the black line denotes a Lorentzian fit. The tone at $\SI{40}{\mega\hertz}$ is due to the probe signal beating against the residual optical carrier. (b) Scaling of the the noise floor, with the electronic noise subtracted, vs local oscillator power. The power law fit gives $p\approx 1.005$, as expected for shot-noise.}
    \label{fig:mechspectrum}
\end{figure}

We choose $\Delta_{\textrm{IF}}/2\pi \approx \SI{20}{\mega\hertz}$ throughout our experiment. A measurement of the mechanical noise spectrum down-shifted to $\Delta_{\textrm{IF}}$ is shown in the red curve of Fig.~\ref{fig:mechspectrum}. We use the residual beating tone between the probe and carrier, at $\SI{40}{\mega\hertz}$, to balance the optical paths in our detection, and to optimize polarization of the local oscillator. 

\begin{figure}
    \includegraphics{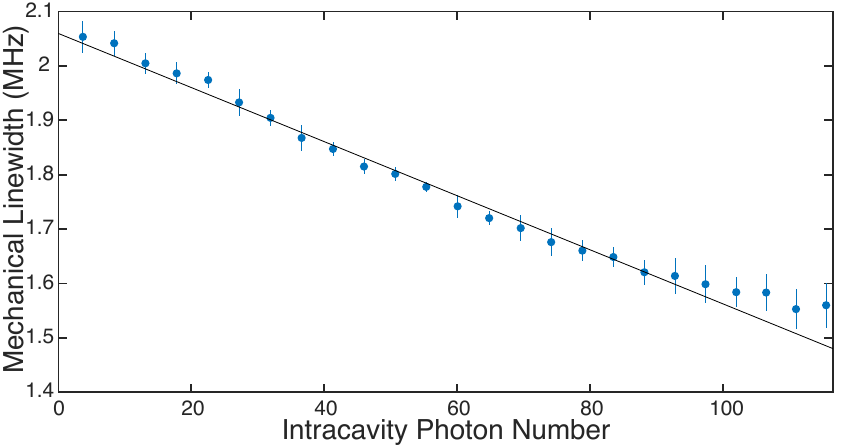}
    \caption{Mechancial linewidth versus intracavity photon number obtained from a laser power sweep. The cavity is pumped on the blue side. From the fit we extract the optomechanical coupling rate: $g_{0}/2\pi = \SI{1.01\pm0.02}{\mega\hertz}$.}
    \label{fig:g0}
\end{figure}

To determine the optomechanical coupling rate we measure mechanical linewidth narrowing, due to backaction, versus laser power. The result is shown in Fig.~\ref{fig:g0}.

\section{Phonon-subtraction Results}
\begin{figure}
    \includegraphics{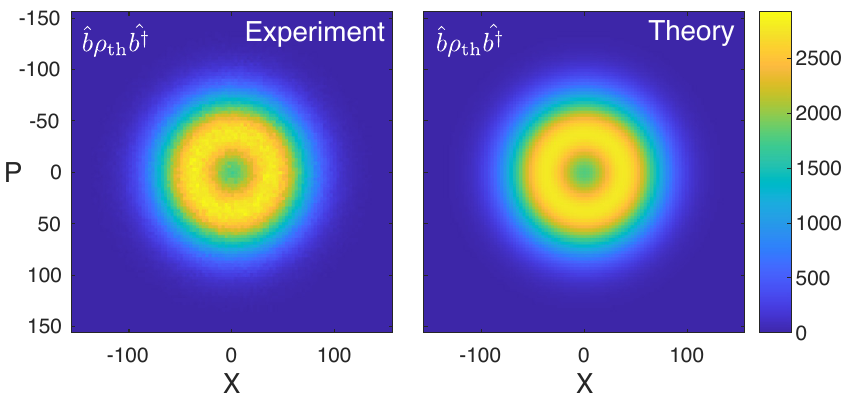}
    \caption{Experimental data and theory fit for quadrature histograms obtained after phonon-subtraction.}
    \label{fig:2D_phon_sub}
\end{figure}

In the main text we described the results of phonon-subtraction using radially binned histograms. In Figure~\ref{fig:2D_phon_sub} we show the two dimensional quadrature histogram for the post-selected experimental data along with a theory fit. 

\begin{figure*}
\includegraphics{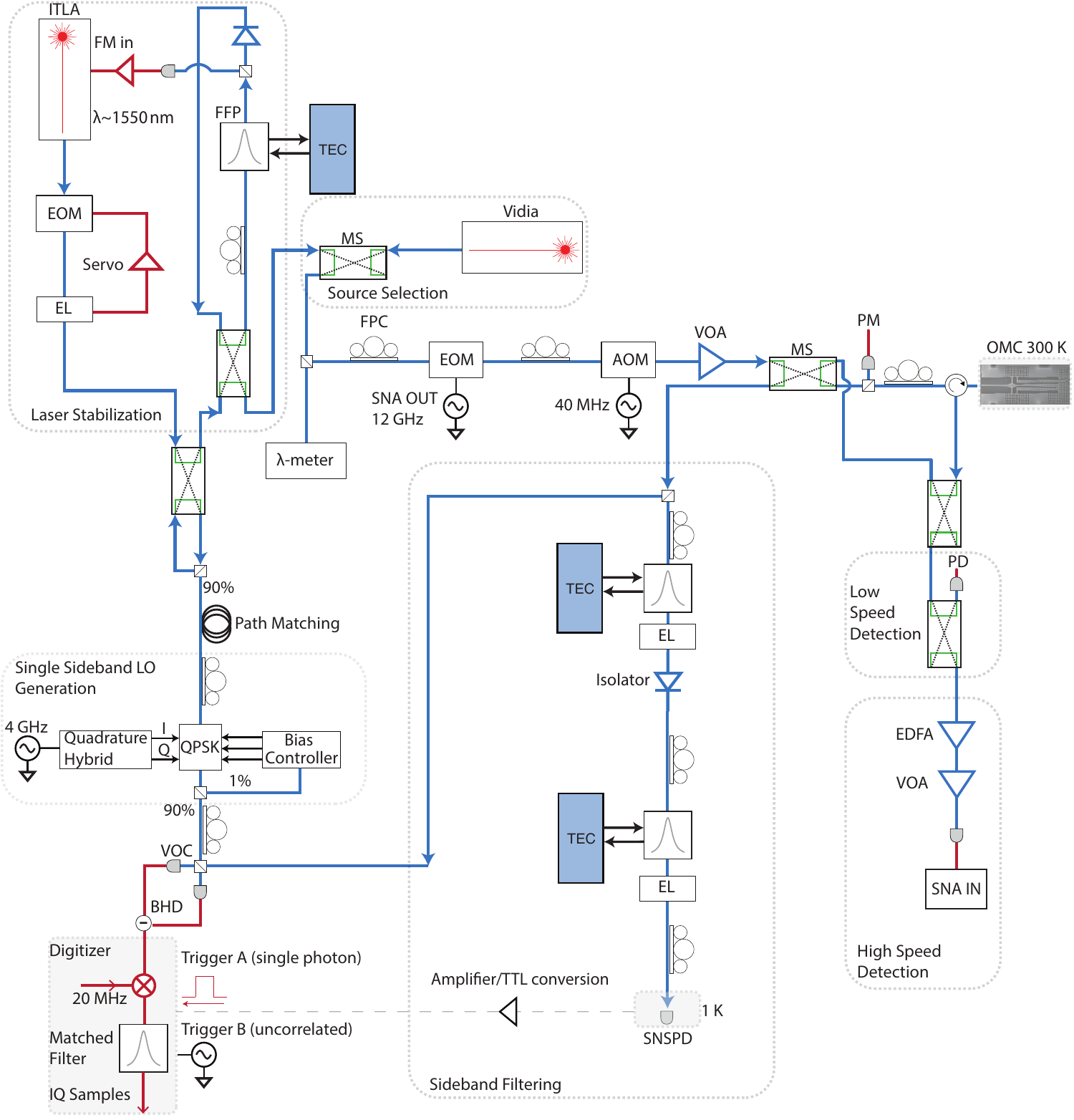}
\caption{Detailed diagram of optomechanical tomography setup. EOM: Electro-optic modulator. FFP: Fiber fabry-perot filter (see main text for details). EL: inline eigenlight power tap. MS: MEMS optical switch (the BAR state is solid green; the CROSS state is dotted black). FPC: Fiber polarization controller. SNA: Scalar network analyzer. VOA: Variable optical attenuator. VOC: Variable optical coupler. QPSK: Quadrature phase shift keying modulator. TEC: Thermo-electric cooler controller. PD: Photodiode. PM: Power meter. EDFA: Erbium doped fiber amplifier. SNSPD: Superconducting nanowire single photon detector. BHD: Balanced heterodyne detection.}
\end{figure*}    

\section{Radial MaxLik Tomography}
\subsection{Definition of Radial POVMs}

The POVM (positive-operator-value-measure) operators that describe heterodyne measurement results are given by: 
\begin{equation}
\label{heterodynePOVM}
\Op{\Pi}{}(\alpha) = \frac{1}{\pi} \ketbra{\alpha}{\alpha}
\end{equation}
for all complex $\alpha$. However, for a measurement apparatus such as ours which is insensitive to phase, we define a set of rotationally symmetric POVMs via phase averaging with  
\begin{equation}
\label{radialsym POVM}
    \frac{\povmRS}{r} = \int_{0}^{2\pi} \Op{\Pi}{}(\alpha) {d\phi}
\end{equation}
where we take $\alpha = r e^{i\phi}$.

The resulting POVM operators corresponding to measurements $|\alpha|=r$ are diagonal with the elements: 
\begin{equation}
    \label{noiseless_radial_povm}
    \bra{n} \povmRS \ket{n} = \frac{2e^{-r^2} r^{2n + 1}}{n !}.
\end{equation}
Note that we have defined $\Pi_{\textrm{RS}}(r)$ such that the following condition required by POVMs is satisfied: 
\begin{equation}
    \int_{0}^{\infty} \povmRS dr = \hat{I}
\end{equation}
where $\hat{I}$ is the identity matrix. 

In a numerical implementation, the POVM must be evaluated on a discrete vector of radius points, and then numerically integrated to generate a probability for each radial bin. This must be done before applying a maximum-likelihood tomography technique, as described in the following sections. Full matrix multiplication is made unnecessary by the choice of radial POVMs, and both the POVMs and the density matrix can be represented by vectors in a numerical implementation. 

Using these POVMs we can write a likelihood function for generating a given dataset in the experiment. Given a set of measurement results $\{p_{i}\}$ where each $p_{i}$ denotes the experimentally measured probability for a result lying in the $i$th radial bin. More precisely, given a set of IQ datapoints $\{s\}$, we construct the radially binned dataset by simply computing: 

\begin{equation}
    p_{i} = P\left(r_{i} < \sqrt{\Re[s]^{2} + \Im[s]^{2}} < r_{i+1}\right)
\end{equation}
where the smallest radial bin edge $r_{0} = 0$ and the largest, $r_{N_{\textrm{bin} - 1}} = R_{\textrm{max}}$ is chosen to capture the largest magnitudes in the data.  

The cost function, $C$ ,to be minimized via the density matrix elements (the optimization parameters) of $\rho$ is the negative log of the likelihood function $\mathcal{L}$ constructed from the POVMs and the experimentally observed data: 

\begin{equation}
\label{loglik}
    C = -\log(\mathcal{L}) = -\sum_{i = 0}^{N_{\textrm{bin}} - 1} p_{i} \log \Tr[\Op{\Pi}{i}{}\rho]
\end{equation}

where we have used the shorthand: 

\begin{equation}
\label{binnedPOVMs}
    \Op{\Pi}{i} = \int_{r_{i}}^{r_{i+1}} \povmRS dr
\end{equation}

In other words, the set $\{\Pi_{i} \}$ are the binned POVMs.

Minimizing Eq. \ref{loglik} via the set of parameters $\{\hat{\rho}_{n,n}\}$ can in principle be done using a variety of methods. In this work, we use the iterative MaxLik algorithm that has found widespread use in continuous variable tomography experiments \cite{lvovsky_iterative_2004,eichler_experimental_2011}.
We form the R matrices using the quantitites defined above as: 

\begin{equation}
    R = \sum_{i} \frac{p_{i}}{\Tr[\Op{\Pi}{i}\rho]} \Op{\Pi}{i}
\end{equation}

The binned POVMs described by Eq. \ref{binnedPOVMs} satisfy the requirement:

\begin{equation}
    \sum_{i} \Op{\Pi}{i} = I
\end{equation}

Of course, an infinite fock space is required for this to be true numerically, but this is not an issue in the radial estimation scenario here because we have a-priori knowledge on the size of the states, and can choose $N_{\textrm{fock}} \gg \bar{n}$. 

Starting from a uniform initial guess: $\rho = \frac{1}{N_{\textrm{fock}}} I$

we update $\rho$ on each step via: 

\begin{equation}
    \rho \rightarrow  \frac{\R\rho \R}{\Tr[\R\rho\R]}
\end{equation}

In practice, a learning rate (step-size) is chosen to dilute R and improve convergence \cite{Rehacek2007}. In this case, R is replaced: $R \rightarrow I + \epsilon R$ where $I$ is the identity matrix, and $\epsilon$ is a learning rate. The exact value of the rate is unimportant, but a value that is too large results in oscillations in the log-likelihood value vs. iteration. We set $\epsilon = 10^{-2}$. 

Finally, we stop the sovler when the trace distance between successive $\rho$ falls below a threhsold, $\varepsilon$: 

\begin{equation}
    \norm{\hat{\rho}_{k} - \hat{\rho}_{k-1}} < \varepsilon
\end{equation}

We obtain good results with $\varepsilon = 10^{-5}$, verified using a known simulated thermal state as input to the solver. 

\subsection{POVMs for states with added noise}
The POVMs given above allow reconstruction of quantum states assuming no additional uncorrelated technical noise has been added to the signal. Following the treatment in \cite{eichler_experimental_2011,eichlerthesis,eichler2012characterizing}, added technical noise can be captured by modifying Eq. \ref{heterodynePOVM} accordingly: 
\begin{equation}
\label{addednoise_heterodyne}
\povmalpha = \frac{1}{\pi} \disp \hat{\rho}_{\text{th}} \dispdagger
\end{equation}
Here $\disp$ is the displacement operator, and  $\hat{\rho}_\textrm{th}$ is defined as in Eq.~\ref{rhothermal} where the bath temperature $\bar{n}$ is replaced by the added effective noise power, in units of phonons. i.e. $\bar{n} = n_{\textrm{added}}$. As described in the main text, in our setup this is estimated by blocking the signal port, and measuring the noise that results. Note in particular that setting $n_{\textrm{added}} = 0$ gives $\rhothermal = \ketbra{0}{0}$ and recovers Eq.~\ref{heterodynePOVM} from Eq. \ref{addednoise_heterodyne}. One can obtain the radially symmetric POVMs in Eq.~\ref{radialsym POVM} by phase averaging over the POVMs in Eq.~\ref{addednoise_heterodyne}. 

In our estimation problem, we require $N_{\textrm{fock}} \gg \bar{n}$. For a $4$~GHz oscillator at room temperature, $\bar{n} \approx 2\cdot 10^{3}$ therefore requiring about $N_{\textrm{fock}} \approx 4\cdot 10^{4}$ to reliably reconstruct the state. Thus, given this enormous Hilbert space dimension, directly evaluating the displacement operators in Eq.~\ref{addednoise_heterodyne} is prohibitive numerically. Since the radial POVMs are diagonal matrices, we do not require the full computation of the displacement operators. In the next section we provide a formula for the on-diagonal elements of Eq.~\ref{addednoise_heterodyne}.  

\subsection{Photon number statistics of a displaced thermal state}

We derive the following results to efficiently generate the POVMs in the previous section numerically. Given: 

\begin{equation}
    \hat{\rho}_{\text{th}} = \left( 1-e^{-\beta} \right) e^{-\beta \hat{n}}, 
\end{equation}

we derive an expression for the diagonal matrix elements of the required POVM:

\begin{equation}
    \left \langle n \middle | \disp \hat{\rho}_{\text{th}} \dispdagger \middle | n \right \rangle = (1-t) t^n e^{\alpha^2 (t-1)} L_n (-A).
\end{equation}

In this equation, we have defined: 
\begin{equation}
    \begin{split}
        t &= e^{-\beta}, \\
        A &= \frac{\alpha^2 (t-1)^2}{t},
    \end{split}
\end{equation}
and the $L_n(x)$ are the Laguerre polynomials. These are given by:
\begin{equation}
    L_n(x)=\sum_{k=0}^n \binom{n}{k}\frac{(-1)^k}{k!} x^k . 
\end{equation}

We set $\alpha = r$ in the above formulas in our numerical implementation.

\end{document}